\documentclass[conference,10pt]{IEEEtran}
\IEEEoverridecommandlockouts
\PassOptionsToPackage{hyphens}{url}
\usepackage{hyperref}
\usepackage{cite}
\usepackage{amsmath,amssymb,amsfonts}
\usepackage{algorithmic}
\usepackage{graphicx}
\usepackage{textcomp}
\usepackage{xcolor}
\def\BibTeX{{\rm B\kern-.05em{\sc i\kern-.025em b}\kern-.08em
    T\kern-.1667em\lower.7ex\hbox{E}\kern-.125emX}}

\newcommand{\ceil}[1]{\left\lceil #1 \right\rceil}    
\renewcommand{\baselinestretch}{0.965}
    
\begin{document}

\title{P$^2$M-DeTrack: \underline{P}rocessing-in-\underline{P}ixel-in-\underline{M}emory for Energy-efficient and Real-Time Multi-Object \underline{De}tection and \underline{Track}ing
}

\author{\IEEEauthorblockN{Gourav Datta$^1$, Souvik Kundu$^1$, Zihan Yin$^1$, Joe Mathai$^2$, Zeyu Liu$^1$, Zixu Wang$^1$, Mulin Tian$^1$, Shunlin Lu$^1$, \\ Ravi Lakkireddy$^1$, Andrew Schmidt$^2$, Wael Abd-Almageed$^2$, Ajey Jacob$^2$, Akhilesh Jaiswal$^{1,2}$, and Peter Beerel$^1$}
\thanks{$^{\dagger}$This work was supported by the DARPA HR$00112190120$ award.}
\it{$^1$University of Southern California, Los Angeles, USA \qquad $^2$Information Sciences Institute, Marina Del Rey, USA} \\
 	Correspondence: \it{gdatta@usc.edu} \\[-1.0ex]}

\maketitle

\begin{abstract}

Today's high resolution, high frame rate cameras in autonomous vehicles generate a large volume of data that needs to be transferred and processed by a downstream processor or machine learning (ML) accelerator to enable intelligent computing tasks, such as multi-object detection and tracking. The massive amount of data transfer incurs significant energy, latency, and bandwidth bottlenecks, which hinders real-time processing. To mitigate this problem, 
we propose an algorithm-hardware co-design framework called \underline{P}rocessing-in-\underline{P}ixel-in-\underline{M}emory-based object \underline{De}tection and \underline{Track}ing (P$^2$M-DeTrack). 
P$^2$M-DeTrack is based on a custom faster R-CNN-based model that is distributed partly inside the pixel array (front-end) and partly in a separate FPGA/ASIC (back-end). The proposed front-end in-pixel processing down-samples the input feature maps significantly with judiciously optimized strided convolution 
and pooling. Compared to a conventional baseline design that transfers frames 
of RGB pixels to the back-end, the resulting P$^2$M-DeTrack designs reduce 
the data bandwidth between sensor and back-end by up 
to $24\times$. 
The designs also reduce the sensor and total energy (obtained from in-house circuit simulations at Globalfoundries 22nm technology node) per frame by $5.7\times$ and $1.14\times$, respectively. Lastly, they reduce the sensing and total frame latency by an estimated $1.7\times$ and $3\times$, respectively. We evaluate our approach on the multi-object object detection (tracking) task of the large-scale BDD100K dataset and observe only a $0.5\%$ reduction in the mean average precision ($0.8\%$ reduction in the identification F1 score) compared to the state-of-the-art.
\end{abstract}
\vspace{2mm}

%%%%%%%%% BODY TEXT
\section{Introduction \& Related Work}
\label{sec:intro}

Artificial intelligence (AI)-enabled video processing presents a challenging problem because the high resolution, high dynamic range, and high frame rates of image sensors generate large amounts of data that must be processed in real-time \cite{sensor1, sensor2}. In particular, the data transmission between the image sensor and the off-chip processing unit leads to significant latency, energy, and bandwidth bottlenecks. This problem is further exacerbated in an autonomous driving scenario, where there are a plethora of other sensors, such as radars and inertial measurement units (IMUs), that also need to transmit data for intelligence processing, including perception and localization \cite{sensors}.

To mitigate this problem, prior works have proposed massively parallel in-memory \cite{li2020timely,datta2022ace}, in-sensor \cite{chen2020pns,memoryless}, and in-pixel computing \cite{Mennel2020UltrafastMV, angizi2022pisa,jaiswal1, jaiswal2,scamp2020eccv,datta2022scireports}, that aims to desegregate sensing, memory, and computation. Some in-sensor computing works implement analog computing in peripheral circuits for mapping AI algorithms \cite{chen2020pns,memoryless} but rely on serial kernel access that incurs significant energy and throughput bottlenecks. On the other hand, in-pixel computing approaches based on emerging technologies \cite{Mennel2020UltrafastMV,angizi2022pisa} promise excellent energy and throughput improvements but are not compatible with the foundry-manufacturing of modern CMOS image sensor (CIS) platforms and hence are difficult to scale. Other solutions that are based on CMOS technology \cite{jaiswal1,jaiswal2,scamp2020eccv} have been limited to toy workloads, such as digit recognition, because they lack the support of multi-channel in-pixel convolutions.

More recently, an in-situ processing-in-pixel-in-memory (P$^2$M) framework has been proposed \cite{datta2022scireports} in which the pixel array implements several deep learning operations, including multi-bit, multi-channel convolution, batch normalization (BN), and ReLU operations. However, this work proposed using non-overlapping strides in the convolutional layer that can lead to a significant drop in accuracy for more complex vision tasks, such as object detection and tracking. In particular, these complex tasks often involve large kernel sizes operating over high-resolution images for which overlapping strides are necessary to avoid missing critical spatial features.
In fact, most of the prior works, including \cite{datta2022scireports}, have been focused on datasets, such as CIFAR-$10$ \cite{song2021reconfigurable}, hyperspectral image recognition \cite{datta2022hsipip} and TinyML applications \cite{NEURIPS2020_ebd9629f}, which have input images
with a significantly lower resolution compared to those in large-scale object detection benchmarks, such as BDD100K \cite{bdd100k}, and do not represent images captured by the existing high resolution (${>}$1M Pixel) image sensors. 

\begin{figure*}
\centering
\includegraphics[width = 0.8\linewidth, height=0.65\linewidth, keepaspectratio]{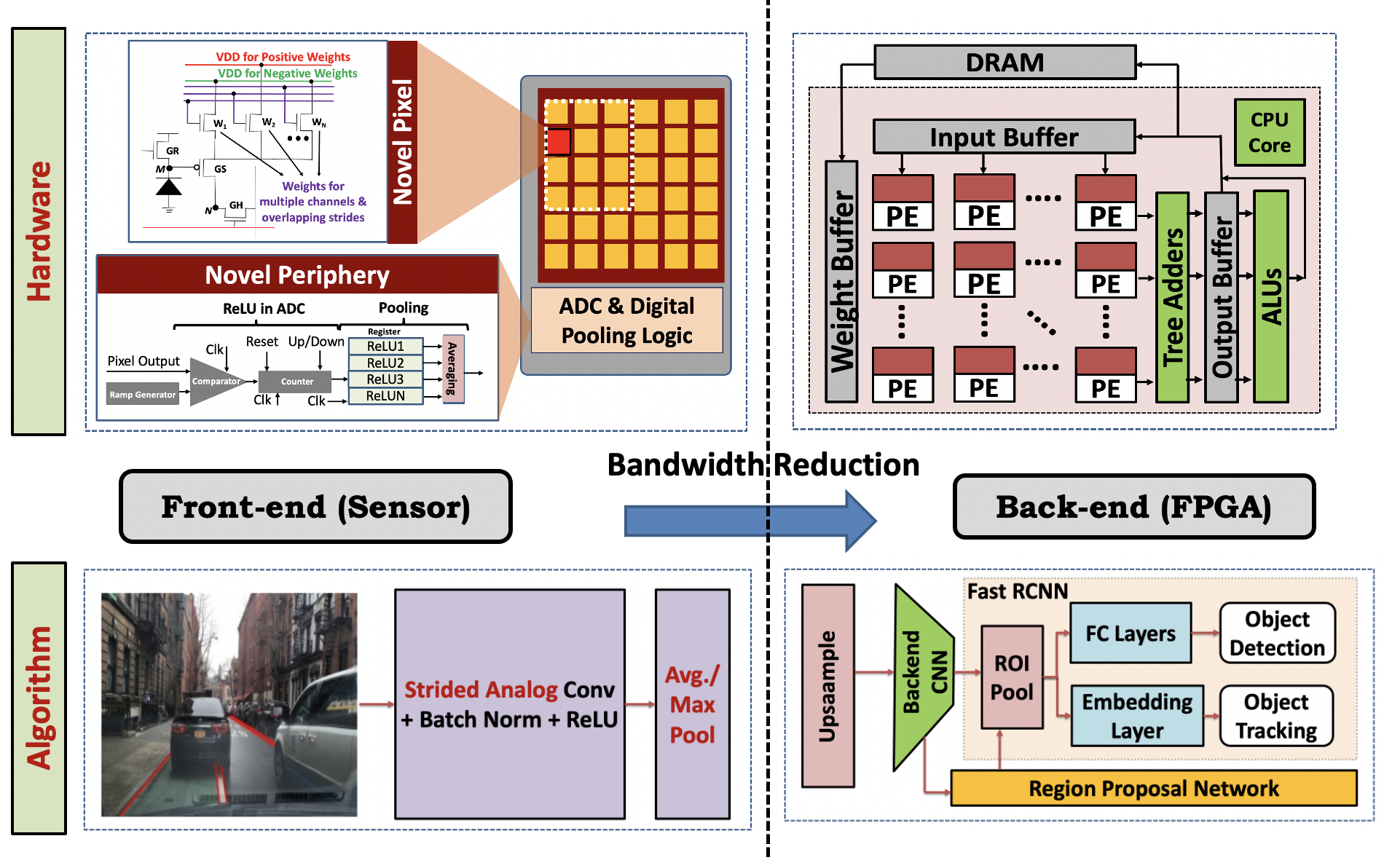}
\caption{Algorithm-hardware co-design framework that enables our proposed P$^2$M approach to optimize both the performance and energy-efficiency of downstream multi-object detection and tracking tasks.}
\label{fig:pip_framework}
\vspace{-3mm}
\end{figure*}

To the best of our knowledge, this work is
the first to show the feasibility of scaling analog processing-in-pixel paradigms, specifically the P$^2$M paradigm, to large-scale complex computer vision applications like real-time multi-object detection and tracking.
As illustrated in Fig. \ref{fig:pip_framework},
our approach termed P$^2$M-DeTrack enables the support for non-overlapping strides that improves task accuracy and max/average pooling that jointly yields up to $24\times$ bandwidth reduction between the image sensor and back-end processing unit. We show how parallelism within P$^2$M-DeTrack mitigates the large latencies induced by non-overlapping strides, thereby meeting the high frames per second (FPS) needed for real-time detection and tracking.
%bandwidth reduction for the object detection
More specifically, we present and evaluate P$^2$M-compatible faster R-CNN-based detection \cite{faster-rcnn} and tracking \cite{qdtrack} models obtained via extensive algorithm-hardware co-design. We choose BDD100K \cite{bdd100k} to validate our approach because it incorporates
geographic, environmental, and weather diversity as well as intentional occlusions, and is the most comprehensive open-source dataset for large-scale object detection and tracking geared towards autonomous driving. 
%The obtained models yield $1.14\times$ and $3\times$  reductions in the total system energy and latency, compared to existing in-sensor solutions.  
%As our models generate compressed outputs from the sensor, 
Our resulting models obtain significant reductions in communication bandwidth, total latency, and energy consumption with negligible drop in detection and tracking accuracies.
%by up to $24\times$, total energy by $1.14\times$, and total latency by $3\times$. 

%1. huge data transfer for video applications, specially autonomous driving because.....
%2. huge demand for edge computing...dont send data to GPUs in cloud
%3. In/near-sensor computing, in-pixel computing...not shown to work for large-scale models before
%4. Our contributions (pooling in sensor, design space exploration for maximal compression, extension to tracking, evaluate in FPGA, energy model)
%5. the paper is structured as....

\section{Preliminaries}

\subsection{P$^2$M: Processing-in-Pixel-in-Memory Paradigm}
The P$^2$M framework proposed in \cite{datta2022scireports} implements multi-bit, multi-channel, non-overlapping strided convolution using memory (weight)-embedded pixels. The weights are represented by the transistor widths \cite{rram}, which modulates the pixel outputs, and the BN and ReLU operations are implemented in the periphery of the pixel array using available on-chip column parallel ADCs in CMOS image sensors. Although the width-encoded weights are fixed during manufacturing, this lack of programmability may be more than an acceptable penalty because the first few layers of any computer vision (CV) model extract high level features that can be common across various visual tasks. Moreover, we can also reconfigure the weights by mapping them to emerging resistive non-volatile memory elements embedded within individual pixels \cite{datta2022scireports}. P$^2$M  leverages the existing on-chip \textit{correlated double sampling} (CDS) circuit in commercial cameras to accumulate both negative and positive weights that are needed to train accurate CV models. Note that the CDS circuit also helps implement the subsequent ReLU operation. Interested readers are referred to \cite{datta2022scireports} for more details of the P$^2$M circuit implementation.

\subsection{Object Detection \& Tracking}

QDTrack is a tracking by detection method \cite{qdtrack} that uses a two stage detector such as faster R-CNN \cite{faster-rcnn} to associate region proposals in temporal neighborhood using contrastive loss. This helps to associate identical objects in the same representation space while pushing dissimilar objects apart. 
QDTrack creates a per frame representation for each object detected which is then used in nearest neighbor search to do association between frames. In this work, we adopt the state-of-the-art (SOTA) QDTrack method with the faster R-CNN architecture for detection to validate our approach.

%or could be mapped to emerging resistive non-volatile memory elements embedded within individual pixels \cite{datta2022scireports}.  Thus, the weights can be either fixed or programmable based on the specific memory technology used for mapping weights within pixel arrays. Interestingly, any 
%By activating multiple pixels simultaneously, the weight modulated outputs of different pixels are summed together in parallel in the analog domain, effectively performing a convolution operation. 

\section{Proposed Methodology}\label{sec:meth}

In this section, we first present our algorithm-hardware co-design framework that enables the support for non-overlapping strides and pooling layers which improves task accuracy. Our framework also enables real-time detection and tracking by exploiting column parallel ADCs for improved analog processing throughput. Lastly, we analytically derive the bandwidth reduction (which leads to energy and latency savings) obtained by P$^2$M-DeTrack.

\subsection{Algorithm-Hardware Co-Design}\label{subsec:algo_HW_co-design}

\textit{Strided Convolution}: In order to implement the strided convolution needed by object detection models, we propose to %increase the number of weight transistors per pixel. 
%Specifically, each pixel needs to be 
embed each pixel with different sets of weight transistors, depending on the value of the stride. 
%However, as illustrated in Eq. \ref{eq:pixel_count}, this approach increases the number of transistors per pixel, which is typically limited by the CIS technology node. 
In particular, as illustrated in Eq. \ref{eq:pixel_count}, the maximum number of required transistors ($N_t$) per pixel is calculated as
\begin{equation}\label{eq:pixel_count}
    N_t=\ceil{\frac{K}{S}}^2*C_o
\end{equation}
\noindent
where $K$, $S$, and $C_o$ are the kernel size, stride, and the number of output channels of the in-pixel layer. Notice that a lower stride value improves detection and tracking accuracy but incurs more transistors and higher required data transmission bandwidth between the sensor and back-end processing unit. %This presents an intricate algorithm-hardware co-design trade-off.  

\textit{Pooling}: The pooling layer in a CNN backbone reduces both the spatial dimensions of a feature map. In order to increase bandwidth reduction, we propose to implement the first pooling layer, following the convolutional, BN, and ReLU layers, inside the P$^2$M chip. To achieve this, we buffer the outputs of the counter (of a single-slope ADC) used to implement the ReLU \cite{datta2022scireports}, and perform an `average' or `max' operation by digital logic in the periphery of the pixel array, as illustrated in Fig. \ref{fig:pip_framework}. Consequently, we can map all the computational aspects of modern CNN layers inside the sensor chip through P$^2$M analog computing.
%, such as multi-bit, multi-channel, strided convolution, linear operation required by BN, non-linear operation required by ReLU and the `average' or `max' operation required by pooling 

\begin{figure}
\centering
\includegraphics[width = 0.99\linewidth]{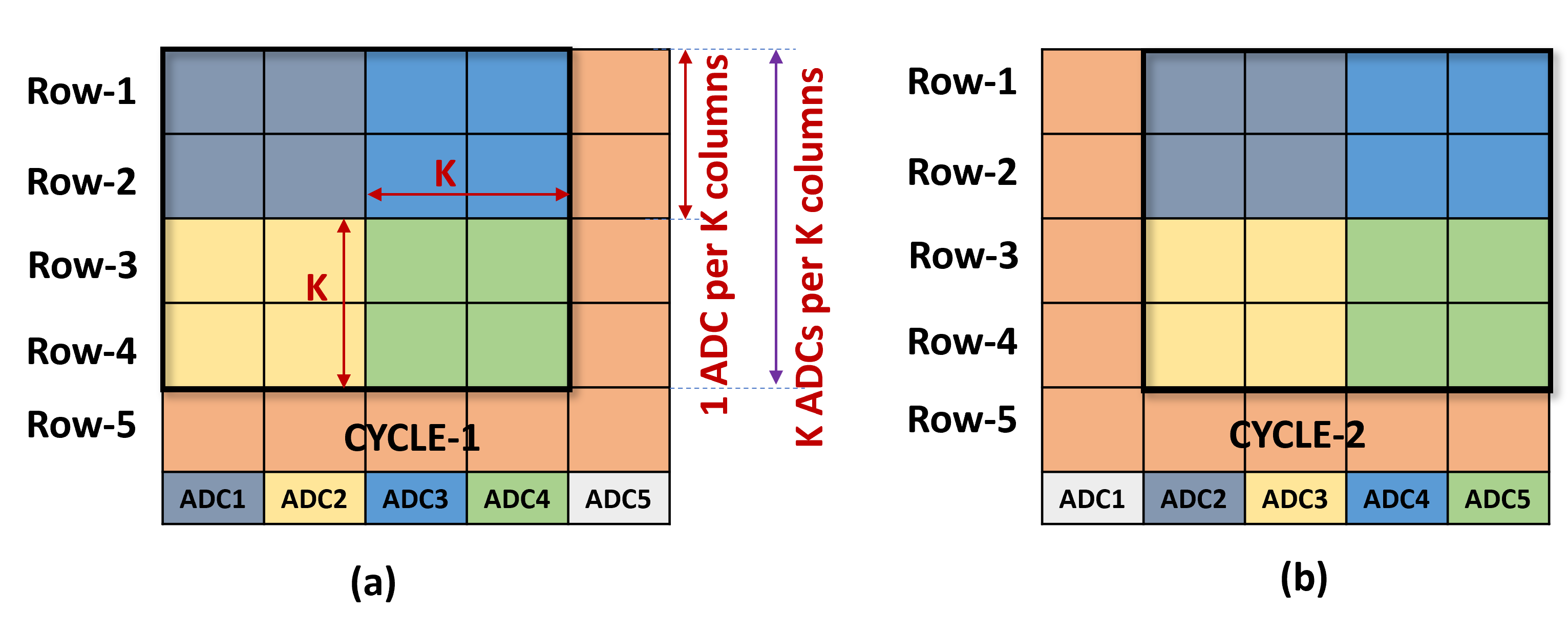}
\caption{For a pixel array of size 5$\times$5, and a $K\times K$ kernel size, $K$ kernel locations can be accessed in parallel on the same column by using $K$-ADCs for $K$-columns by ensuring the resulting kernels do not overlap. (a) and (b) show how to handle a horizontal striding of a stride of 1, where two clock cycles are needed to generate the output activation associated with all horizontal kernel locations. Handling a overlapping vertical stride of 1 is not shown, but can be processed similarly.}
%Note, vertical striding (not shown in the figure) would occur in a similar fashion .}
\label{fig:adc_parallel}
\vspace{-1mm}
\end{figure}

\textit{P$^2$M-DeTrack Parallelism}: Enabling P$^2$M-DeTrack computations inside the pixel array opens up a new opportunity to improve in-pixel parallelism for a given channel. Consider Fig. \ref{fig:adc_parallel} that shows a pixel array of size 5$\times$5 and 2$\times$2 kernel with a stride of 1.
Operating on Row-1 and Row-2 across two cycles (Fig. \ref{fig:adc_parallel}(a) and Fig. \ref{fig:adc_parallel}(b)), all the output activations corresponding to the horizontal striding can be generated. Note, for each kernel one ADC in the periphery, represented with same color coding as the corresponding kernel, is activated to convert the analog convolution operation into digital activations. More generally, this implies for each kernel of size $K\times K$ only one among $K$-ADCs is being utilized, leaving ($K$-1) ADCs idle (not used) per $K$ columns. Fig. \ref{fig:adc_parallel} presents a method to exploit such idle ADCs to improve processing parallelism. Specifically, for each column, $K$ kernels can be activated, simultaneously. $K$ ADCs on $K$ columns would be used to convert the analog convolution output to digital activation. Assuming a stride of $S$, $\ceil{\frac{K}{S}}$ clock cycles are required to process each row, where $\ceil{.}$ denotes the ceiling function. On the other hand, using $K$ ADCs in parallel for each kernel, implies that only $\ceil{\frac{H}{K}}$ cycles are required to process each column, where $H$ is the height of the output activation map (which equals the height of the input activation map divided by $S$). Consequently, the number of clock cycles required to yield the complete digital output activation map for a given channel is 
\begin{align}\label{eq:adc_cycles}
N_c=\ceil{\frac{H}{K}}\cdot\ceil{\frac{K}{S}}    
\end{align}
In this way, P$^2$M-DeTrack enables row and column parallelism, transforming the pixel array into multiple analog parallel processing blocks.

\textit{Discussion:} The parallelism in P$^2$M-DeTrack within a channel is limited by 
the amount of overlapping of the kernels. More
precisely, it takes $\ceil{\frac{K}{S}}$ cycles to execute each pixel row. In addition, the different channels must be executed on the pixel array sequentially. Moreover, the number of required weight transistors per pixel is dependent on the kernel size, stride, and number of channels which is
limited by area and technology constraints  \cite{datta2022scireports}.
These constraints motivate a circuit-algorithm co-design that minimizes the number 
of channels and maximizes the stride $S$ subject to FPS requirements and achieving close to state-of-the-art (SOTA) accuracy. Finally, to
achieve these accuracy goals, we must consider the non-linearities of the analog circuitry in the training process \cite{datta2022scireports}.
%and the area limitations associated with the pixel transistors 
%However, P$^2$M \cite{datta2022scireports} focused on TinyML workloadsfewer, and hence, could restrict the number of output channels than needed for accurate object detection 
%significantly which can impact accuracy of more complex vision tasks.
%. In contrast, we target more complex tasks which not only demands more output channels for fine-grained features, but also require overlapping strides (implementation detail explained above). 
%Both these modifications hurt parallelism and bandwidth reduction\footnote{Note that we partially solve the bandwidth reduction problem by implementing the pooling layer inside the P$^2$M chip.}. 
%
%This optimization problem sets up a circuit-algorithm trade-off that we navigate via a holistic design space exploration.
%We search over kernel and stride sizes and the number of output channels, maximizing bandwidth reduction and minimizing the number of weights per pixel while maintaining close to SOTA multi-object detection and tracking performance. 
In particular, we focus on maximizing the mean average precision (mAP) corresponding to an intersection of union (IoU) averaged uniformly from 0.5 to 0.95 with a step size of 0.05, and mean identification F1 score (IDF1) \cite{qdtrack} for object detection and tracking, respectively.
%wherein the 
%backbone CNN has to be optimized for having
%larger kernel sizes (that increases the concurrent activation of more pixels, helping parallelism), 
%larger channels (that increases the number of transistors per pixel) and overlapping strides (to
%reduce the dimensionality in the downstream CNN layers, thereby reducing the number of multiply-and-adds and peak memory
%usage),while maintaining close to state-of-the-art
%detection and tracking performance and taking into account the non-idealities associated with analog convolution operation \cite{}. %Also, decreasing number of channels decreases the number of weight transistors embedded within each pixel (each pixel has weight transistors
%equal to the number of channels in the output feature map), improving area and power consumption. 
%Furthermore, the resulting
%smaller output activation map (due to reduced number of channels, and larger kernel sizes with non-overlapping strides) reduces
%the energy incurred in transmission of data from the CIS to the downstream CNN processing unit and the number of floating
%point operations (and consequently, energy consumption) in downstream layers, as shown in Section \ref{}.

\subsection{Bandwidth Reduction}

%We perform a thorough design space exploration to compress the sensor outputs subject to negligible reduction in the object detection and tracking performance.

To quantify the bandwidth reduction (BR) by the P$^2$M-implemented layers, let the number of elements in the RGB input image be $I$ and in the output activation map after the pooling layer be $O$. Then, $BR$ can be estimated as
\begin{equation}
BR=\left(\frac{O}{I}\right)\left(\frac{4}{3}\right)\left(\frac{12}{N_b}\right)\label{eq:DR_1}
\end{equation}
\noindent
%where $N_b$ denotes the bit-precision of the convolutional output feature map.
Here, the factor $\left(\frac{4}{3}\right)$ denotes the demosaicing operation \cite{demosaicing}, which converts the Bayer’s pattern of RGGB pixels to RGB pixels by either ignoring the additional green pixel or designing the circuit to effectively take the average of the photo-diode currents coming from the green pixels. The factor $\frac{12}{N_b}$ denotes the ratio of the bit-precision between the image pixels captured by the sensor (pixels typically have a bit-depth of $12$ \cite{onsemi:AR0135AT}) and the quantized output of the first convolutional layer denoted as $N_b$. Let us now substitute
\begin{equation}
O={\left(\frac{\left(\frac{i{-}K'{+}2*D'}{S'}{+}1\right){-}K{+}2{*}D}{S}{+}1\right)^2{*}C_{o}}, \quad I=i^2{*}3 \notag
\end{equation}
\noindent
into Eq. \ref{eq:DR_1}, where $i$ denotes the spatial dimension of the input image and $C_o$ denotes the number of output channels of the first convolutional layer. While $K$, $D$, $S$ denote the kernel size, padding and stride of the in-pixel convolutional layer, respectively, similar notations with the prime superscript $'$ denote the same hyperparameters of the pooling layer. We report the BR obtained by some of the these hyperparameter combinations (selected based on the design space exploration approach discussed above) in Table \ref{tab:mot_compression}.   %, that does not lead to a significant degradation in the detection and tracking performance. 

\renewcommand{\arraystretch}{1.4}
\begin{table}
\begin{center}
\scriptsize\addtolength{\tabcolsep}{-1.2pt}
\begin{tabular}{|c|c|c|c|c|}
\hline
Convolutional & Pooling & ReLU output & Transistors/ & Bandwidth \\
stride ($S$) & stride ($S'$) & bit-width ($N_b$) & pixel ($N_t$) & Reduction (BR) \\
\hline
2 & 2 & 8 & 256 & 6  \\
 %& CNN-32H & & & & \\
\hline
4 & 2 & 8 & 64 & 24 \\
 %& CNN-32H & & & & \\
\hline
6 & - & 8 & 64 & 13.5\\
\hline
\end{tabular}
\end{center}
%\vspace{-3mm}
\caption{Values of different training hyperparameters that do not lead to significant drop in the detection and tracking performance, and the corresponding bandwidth reduction (BR) and the maximum transistor count per pixel ($N_t$).} %We represent $s$ as the stride in ($H_l^i$, $W_l^i$, $D_l^i$).}
\label{tab:mot_compression}
\vspace{-3mm}
\end{table}

\section{Simulation Results}

In this section, we first introduce the implementation details and present our detection and tracking results on the BDD100K dataset. We then provide the energy consumption and FPS improvements obtained by our approach.

\subsection{Experimental Setup}

We evaluate P$^2$M-DeTrack on BDD100K detection and tracking datasets with 12 and 10 object categories, respectively. While the detection set includes 70,000 images for training, and 10,000 images for validation, the tracking dataset includes 1,400 videos (278K images) for training and 200 videos (40K images). All these images have a resolution of 1.06M Pixel. Note that we report results on the validation dataset. The images in the tracking set are annotated per $5$ FPS with a $30$ FPS video frame rate\footnote{We observe that dropping every other frame does not lead to any degradation in the tracking IDF1 score for the BDD100K dataset.}. We use the 2x training schedule consisting of 24 epochs and the 1x training schedule consisting of 12 epochs for our object detection and tracking experiments respectively. While our detection models are trained on 2 GPUs, each with a batch size of 2, our tracking models use 4 GPUs, each with a batch size 10. For both these experiments, we use the SGD optimizer with an initial learning rate of 0.0025.

For a fair comparison, all results are produced using the MMDetection framework \cite{mmdetection} and the standard COCO-style evaluation protocol. In addition, we use ImageNet-pretrained ResNet-50 as the backbone to all our models and fine-tune the learning rate schedules of each model to obtain the best detection mAP. All our proposed models process the first four layers (convolutional+BN+ReLU+pooling) inside the sensors and the remaining layers in the back-end, while the baseline models process the entire neural network in the back-end.

\renewcommand{\arraystretch}{1.4}
\begin{table}
\begin{center}
\scriptsize\addtolength{\tabcolsep}{-2.2pt}
\begin{tabular}{|c|c|c|c|c|c|c|}
\hline
Model & $S$ & Pool & $S'$ & mAP (\%) & mAP (\%) & mAP (\%) \\
 & & type & & (IoU=0.5) & (IoU=0.75) & (IoU=0.5-0.95) \\
 \hline
Faster R-CNN \cite{faster-rcnn} & 2 & Max & 2 & 57.5 & 33.2 & 33.7 \\
\hline
P$^2$M-DeTrack & 2 & Max & 2 & 56.5 & 32.9 & 33.2 \\
\hline
P$^2$M-DeTrack & 6 & - & - & 55.5 & 31.7 & 32.3 \\
\hline
P$^2$M-DeTrack & 4 & Avg & 2 & 53.5 & 30.3 & 31.1 \\
\hline
\end{tabular}
\end{center}
%\vspace{-3mm}
\caption{mAP values corresponding to three standard IoUs of the baseline faster R-CNN and our P$^2$M-DeTrack models for different in-pixel convolutional and pooling strides.} %We epresent $s$ as the stride in ($H_l^i$, $W_l^i$, $D_l^i$).}
\label{tab:detection_map}
\vspace{-5mm}
\end{table}

\subsection{Bandwidth vs. number of transistors}

As summarized in Table \ref{tab:mot_compression}, we evaluated several model configurations that explore the trade-off between the bandwidth reduction and the number of weight transistors per pixel. With a kernel size of 7, a stride of 4 in the convolutional layer and a stride of 2 in pooling layer leads to a BR of 24$\times$, but requires a maximum of 64 weight transistors per pixel. 
%which may be infeasible with the foundry-compatible technology nodes. 
A convolutional stride of 6 with no pooling leads to a BR of 13.5$\times$, also with a maximum of 64 weight transistors per pixel\footnote{Note that fewer pixels need 64 weight transistors for a stride of 6 than for a stride of 4.}. Note, using chip stacking in which weight transistors are integrated vertically on a stacked chip through metal-to-metal fusion bonding \cite{gao2019chip} or through-silicon-vias (TSVs) \cite{motoyoshi2009through}, minimal to no increase in pixel area is expected because of the dense metal-pitch (MP) and contacted poly-pitch (CPP) \cite{gupta2020high} of advanced technology nodes and the relatively large sizes of underlying pixel arrays. 

In addition, to improve the accuracy of downsampled output feature maps of the sensor, the feature maps are upsampled in the back-end processing unit, as shown in Fig. \ref{fig:pip_framework}. In particular, the spatial dimension at the input of the back-end is chosen to be no less than $\left(\frac{1}{4}\right)^{th}$ of the image dimension. %The excellent bandwidth reduction obtained by our proposed P$^2$M-DeTrack validates the efficacy of our proposal. 
In general, the data generated from sensor has to be transferred through costly wired or wireless channels, which could form the key energy bottleneck for the overall system, particularly if the physical distance between the front- and back-end is large.
%All these hyperparameters, along with $N_b$, are obtained via the design space exploration approach explained in Section XX. We show their values in Table \ref{tab:notations}, and substitute them in Eq. \ref{eq:DR_1} to obtain a DR of upto $24\times$.
%\vspace{-2mm}
\subsection{Detection Results}

\begin{figure}
\centering
\includegraphics[width = 0.99\linewidth]{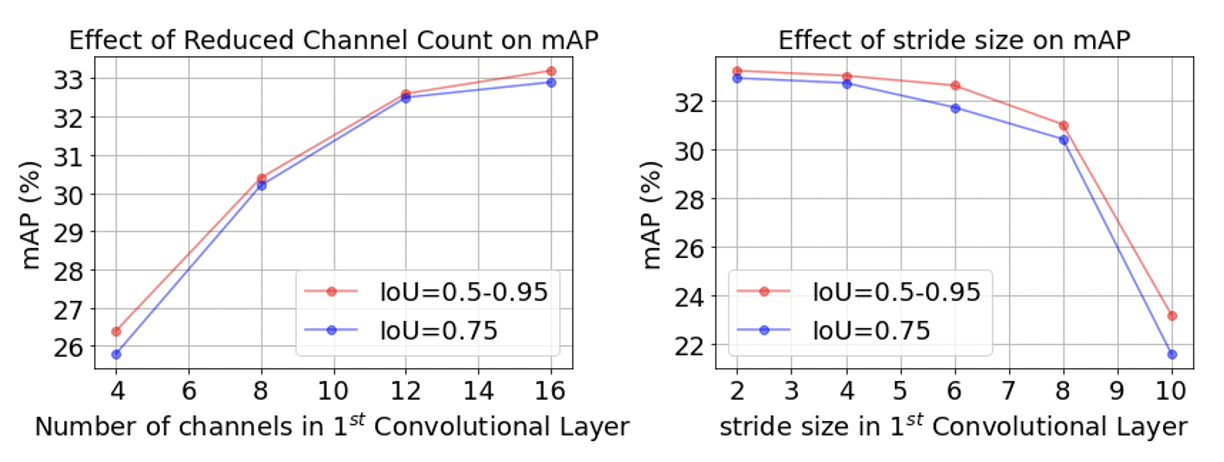}
\caption{Comparison of object detection mAP for various stride and channel counts of the in-pixel convolutional layer.}
\label{fig:ablation}
\vspace{-3mm}
\end{figure}

We developed multiple detection and tracking models with different strides in the convolutional and pooling layers (see Table \ref{tab:mot_compression}) as shown in Table \ref{tab:detection_map}.
The evaluated models in Table \ref{tab:mot_compression} yield mAPs for IoU=0.5:0.95 within 1\% of the baseline model, as detailed in Table \ref{tab:detection_map}. 
We also ablate over the number of channels and strides in Fig. \ref{fig:ablation} to understand the limits of bandwidth reduction. The figure shows that the mAP drops beyond 1\% as we decrease the number of channels below 16. A similar significant accuracy drop is observed for strides above 6. Note that we also increase the kernel size (i.e., make the kernel size equal to stride) to support strides of 8 and 12.

\subsection{Tracking Results}

The detailed breakdown of our tracking results on the BDD100K multi-object tracking (MOT) validation set are shown in Table \ref{tab:hsi_compression}.
Our P$^2$M-DeTrack models achieve 49.7\% mIDF1 (stride 2, pool 2) and 49.1\% mIDF1 (stride 6, pool 2, followed by upsampling), both of which are comparable to the SOTA 50.8\% mIDF1 obtained by QDTrack \cite{qdtrack}.

\renewcommand{\arraystretch}{1.4}
\begin{table}
% results are generated using ../work_dirs/quant_stem_16_7x7_6_qdtrack-frcnn_r50_fpn_12e_bdd100k/epoch_36.pth
% results are generated using ../work_dirs/quant_stem_16_7x7_2_qdtrack-frcnn_r50_fpn_12e_bdd100k/epoch_31.pth
\begin{center}
\scriptsize\addtolength{\tabcolsep}{-5.0pt}
\begin{tabular}{|c|c|c|c|c|c|c|c|c|c|c|}
\hline
Model & $S$ & $S'$ & mMOTA & mIDF1 & MOTA & IDF1 & FN & FP & ID Sw. &mAP \\
 \hline
QDTrack \cite{qdtrack} & - & - & 36.6 & 50.8 & 63.5 & 71.5 & 108614 & 46621 & 6262 & 32.6\\
\hline
P$^2$M-DeTrack & 2 & 2 & 34.0 & 49.7 & 62.4 & 70.7 & 119256 & 41466 & 5964 & 30.8 \\
\hline
P$^2$M-DeTrack & 6 & - & 34.2 & 49.1 & 61.8 & 70.3 & 122691 & 40480 & 5999 & 30.3 \\
\hline
\end{tabular}
\end{center}
%\vspace{-3mm}
\caption{Multi-object tracking (MoT) metrices of the baseline faster R-CNN-based QDTrack and P$^2$M-DeTrack models for different in-pixel convolutional and pooling strides.} %We epresent $s$ as the stride in ($H_l^i$, $W_l^i$, $D_l^i$).}
\label{tab:hsi_compression}
\vspace{-3mm}
\end{table}

\vspace{-1mm}
\subsection{Reduction in Energy Consumption}

As described in \cite{datta2022scireports}, the total energy consumption for both P$^2$M-DeTrack and conventional baseline models can be partitioned into three major components, sensor ($E_{sens}$), sensor-to-SoC communication ($E_{com}$), and SoC energy ($E_{soc}$) as captured in Eq. \ref{eq:e_tot}. Note that $E_{sens}$ is the sum of the pixel array energy\footnote{The pixel array energy is equal to the image read-out energy for the baseline models and in-pixel convolution energy for custom models.} and the ADC energy; both are obtained from circuit simulations using commercial Globalfoundries 22nm FD-SOI technology node. $E_{com}$ is computed from $e_{com}$ as shown in Eq. (4), and its value shown in Table \ref{tab:delay_variable_values} is obtained from \cite{kodukula2020sensors}. On the other hand, $E_{soc}$ in primarily dominated by multiply-and-add (MAdd) energy of the models used for detection ($E_{mac}$).
\begin{align}
E_{tot} \approx \underbrace{(e_{pix}+e_{adc})*N_{pix}}_{E_{sens}} + \underbrace{e_{com}*N_{pix}}_{E_{com}} + \underbrace{e_{mac}*N_{mac}}_{E_{mac}}
\label{eq:e_tot}
\end{align}
Fig. \ref{fig:p2m_base_en_compare} shows the normalized energy comparison between P$^2$M-DeTrack and baseline detection models. In particular, the in-pixel architecture is estimated to reduce the sensing energy by up to $5.7\times$ with a total energy improvement of $1.14\times$. Note, in Eq. \ref{eq:e_tot} $N_{pix}$ and $N_{mac}$ corresponds to the total number of pixels needed to be processed in-pixel and total MAC operations \cite{kundu2020pre} for the back-end model, respectively. We note that the dominating back-end can be often over-parameterized and thus various off-the-shelf model compression \cite{kundu2021dnr, ren2019admm} can be deployed to further reduce the back-end computation cost. This, in turn, can further reduce the total energy. 

%\begin{comment}

\subsection{Reduction in Delay}
The delay associated with the detection pipeline can be partitioned in to four components, sensing delay ($\mathcal{T}_{sens}$), ADC delay ($\mathcal{T}_{adc}$), communication delay ($\mathcal{T}_{com}$), and backend delay ($\mathcal{T}_{back}$) as shown below 
\begin{align}
    \mathcal{T}_{delay} \approx \mathcal{T}_{sens} + \mathcal{T}_{adc} + \mathcal{T}_{comm} + \mathcal{T}_{back}
\end{align}
\noindent
We use the per-pixel communication delay from \cite{Gan2020LowLatencyPC} to estimate the delay required to send the output activation maps from sensor to SoC ($\mathcal{T}_{com}$). 
While $\mathcal{T}_{sens}$ is primarily governed by the sensor read delay, $\mathcal{T}_{adc}$ is associated with the total number of ADC clock cycles required to generate the first layer output activation map, dictated by our P$^2$M-DeTrack parallelism approach discussed in Section \ref{subsec:algo_HW_co-design}. Note that both of these values are computed from our circuit simulations. %, while the ADC delay is associated with the total ADC operations while generating the pixels for first layer output feature map. 

\renewcommand{\arraystretch}{1.3}
\begin{table}
\begin{center}
\scriptsize\addtolength{\tabcolsep}{-2pt}
\begin{tabular}{|c|c|c|}
\hline
 Notation & Description & Value \\ 
 \hline
 $e_{pix}$ & Sensing energy/\#pixel & 148 pJ  (P$^2$M) \\
 {} & {} & 312 pJ  (baseline)\\
 \hline
  $e_{adc}$ & ADC energy/\#pixel & 41.9 pJ  (P$^2$M) \\
 {} & {} & 86.14 pJ  (baseline)\\
 \hline
   $e_{com}$ & SoC comm. energy/\#pixel & 900 pJ\\
  \hline
   $e_{mac}$ & Madds energy in $22nm$ technology & 1.568 pJ\\
  \hline
\end{tabular}
\end{center}
\caption{The description and values of the notations used for the baseline and P$^2$M-DeTrack energy computation \cite{datta2022scireports}.}
\label{tab:delay_variable_values}
\vspace{-4mm}
\end{table}

$\mathcal{T}_{back}$ was computed by the Xilinx FINN-based \cite{finn} FPGA simulation framework targetting the Xilinx Alveo U250 platform, which was modified to include the QSFP+ high-speed transceiver for streaming purposes. Our framework utilizes High-Level Synthesis (HLS) to translate the developed PyTorch model into a synthesizable RTL to generate a bitstream for the target Alveo U250 device. The HLS flow evaluates the network layers, splits each layer into submodules, and applies transformations to aid in the HLS conversion process.  These transforms are intended to optimize the HLS process by enabling FPGA-friendly quantization, pipeline the FIFO insertions, and implementations for efficient memory (BRAM, URAM and LUTRAM) and processing (DSP) for all layers. The resulting model is stitched back together in RTL submodules and run through synthesis, place-and-route, leading to the final bitstream generation.  This process allows for refining individual modules and stages with HLS and analyzing latency and resource utilization for the system bottlenecks. 

\begin{figure}
\centering
\includegraphics[width = 1.0\linewidth]{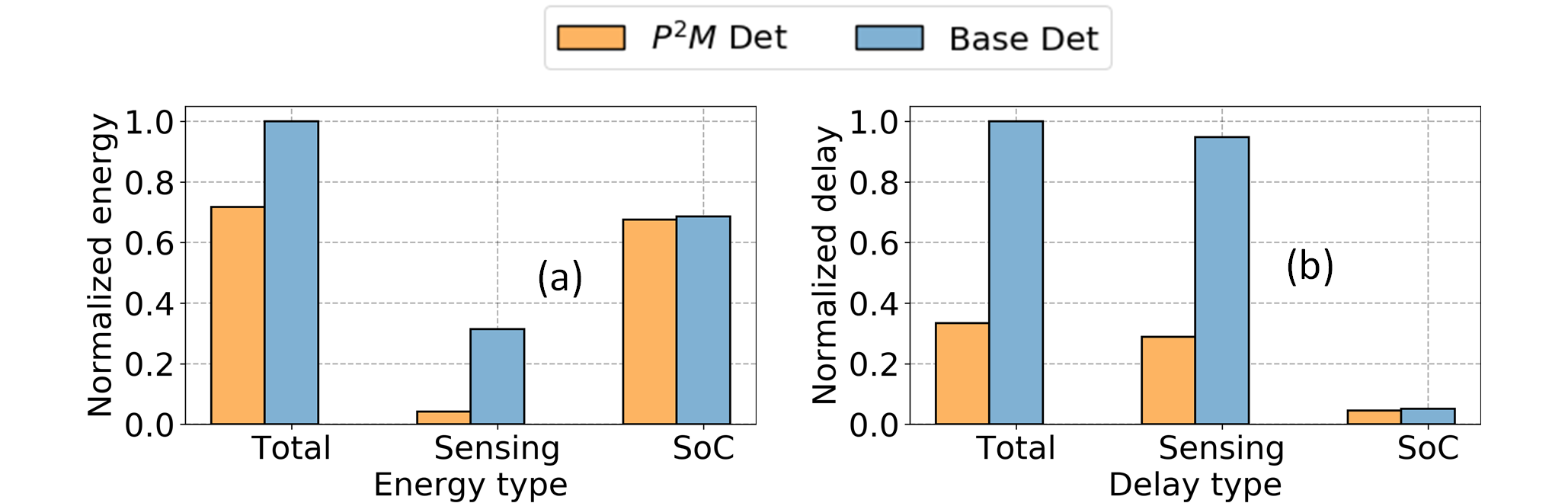}
%\vspace{-1mm}
\caption{Comparison between (a) normalized energy and (b) normalized delay between our proposed P$^2$M and a conventional baseline detection model.}
\label{fig:p2m_base_en_compare}
\vspace{-2mm}
\end{figure}

The resulting inference latency through the system implementing only the CNN backbone is 15.5 ms.  The resource utilization is ~60\% LUTs, 30\% Flip-Flops, 13\% DSPs, 60\% BRAMs, 10\% LUTRAMs and 2\% URAMs.  The power utilization is 70W based on default switching models as reported by the Xilinx Vivido tools.  Augmenting the baseline ResNet-50 model to our proposed custom network with the first convolutional, BN, ReLU, and pooling layers removed, results in a $1.15 \times$ improvement in latency, down to $13.5$ ms. As shown in Fig. \ref{fig:p2m_base_en_compare}, the total delay improvement for in-pixel architecture is estimated to be up to $\mathord{\sim}3\times$. In particular, the proposed P$^2$M-DeTrack model can yield an improved $\mathord{\sim}17$ FPS compared to the baseline's $\mathord{\sim}9.6$ FPS, which may be acceptable for real-time detection and tracking in an autonomous driving scenario. Note that the FPS can be further improved by using faster pixels or by designing and leveraging more than one ADC per column. For example, using 2 ADCs per column can double the FPS at the cost of only modest increase in area.

%The energy-delay product (EDP) improvement in sensing and communication are estimated to be xx and xx respectively.

\section{Conclusions and Discussions}
This work proposes P$^2$M-DeTrack, a processing-in-pixel-in-memory-based algorithm-hardware co-design framework that can 
reduce the EDP cost of a complex object detection 
and tracking pipeline by up to $3.42\times$ ($3\times$ reduction in latency for a $14\%$ reduction in energy) compared to existing in-sensor processing approaches. Our hardware modifications can be readily integrated into the foundry-manufacturable CMOS image sensor platforms, and our hardware-inspired algorithmic modifications are shown to yield negligible performance drop in multi-object detection and tracking. 
To the best of our knowledge, we are the first to incorporate all the computational aspects of modern CNN layers inside CMOS image pixels, and show the feasibility of in-pixel computing for complex vision tasks on high-resolution (${>}$1Mpixel) datasets, including real-time multi-object detection and tracking. Our future work includes pruning and quantizing the faster R-CNN back-end to further optimize the system FPS and energy consumption. We also plan to incorporate the region proposal network of the faster R-CNN architecture in our FPGA simulation and analytical framework, which would further improve our energy and delay estimation models.

It is important to note that the energy metric presented in this work is based on the assumption that the front-end and back-end chips are closely located on the same printed-circuit board \cite{kodukula2020sensors}. In general, the front-end and back-end sensors could be separated by large distances, necessitating long energy-expensive wired or wireless data transfer (which is common for the case of sensor-fusion and swarm intelligence applications). In such cases the overall energy improvement would approach the bandwidth reduction (up to 24$\times$) obtained from the proposed P$^2$M-DeTrack scheme.

{\small
\bibliographystyle{ieee_fullname}
\bibliography{egbib}
}

\end{document}